\documentclass[3p,times]{elsarticle}

\usepackage{ecrc}

\usepackage{wrapfig}

\usepackage{epstopdf}

\volume{00}

\firstpage{1}

\journalname{Nuclear Physics A}

\runauth{H.~Paukkunen et.al.}

\jid{nupha}

\jnltitlelogo{Nuclear Physics A}

\usepackage{graphicx}
\usepackage{amsmath,amssymb}

\begin{document}

\begin{frontmatter}



\title{Dijets in p+Pb collisions and their quantitative constraints for nuclear PDFs}



\author[label1,label2]{Hannu Paukkunen}
\ead{hannu.paukkunen@jyu.fi}
\author[label1,label2]{Kari J. Eskola}
\ead{kari.eskola@jyu.fi}
\author[label3]{Carlos Salgado}
\ead{carlos.salgado@usc.es}

\address[label1]{Department of Physics, University of Jyv\"askyl\"a, P.O. Box 35, FI-40014 University of Jyv\"askyl\"a, Finland}
\address[label2]{Helsinki Institute of Physics, University of Helsinki, P.O. Box 64, FI-00014, Finland}
\address[label3]{Departamento de F\'\i sica de Part\'\i culas and IGFAE, Universidade de Santiago de
Compostela, E-15782 Galicia, Spain}

\begin{abstract}
We present a perturbative QCD analysis concerning the production of high-pT dijets in p+Pb collisions
at the LHC. The next-to-leading order corrections, scale variations and free-proton PDF uncertainties
are found to have only a relatively small influence on the normalized dijet rapidity distributions. 
Interestingly, however, these novel observables prove to retain substantial sensitivity to the nuclear
effects in the PDFs. Especially, they serve as a more robust probe of the nuclear gluon densities at
$x>0.01$, than e.g. the inclusive hadron production. We confront our calculations with the recent data 
by the CMS collaboration. These preliminary data lend striking support to the gluon antishadowing similar
to that in the EPS09 nuclear PDFs.
\end{abstract}


\end{frontmatter}

\section{Introduction}
\label{intro}

\begin{wrapfigure}{r}{0.40\textwidth}
\vspace{-0.9cm}
\centerline{\includegraphics[width=0.40\textwidth]{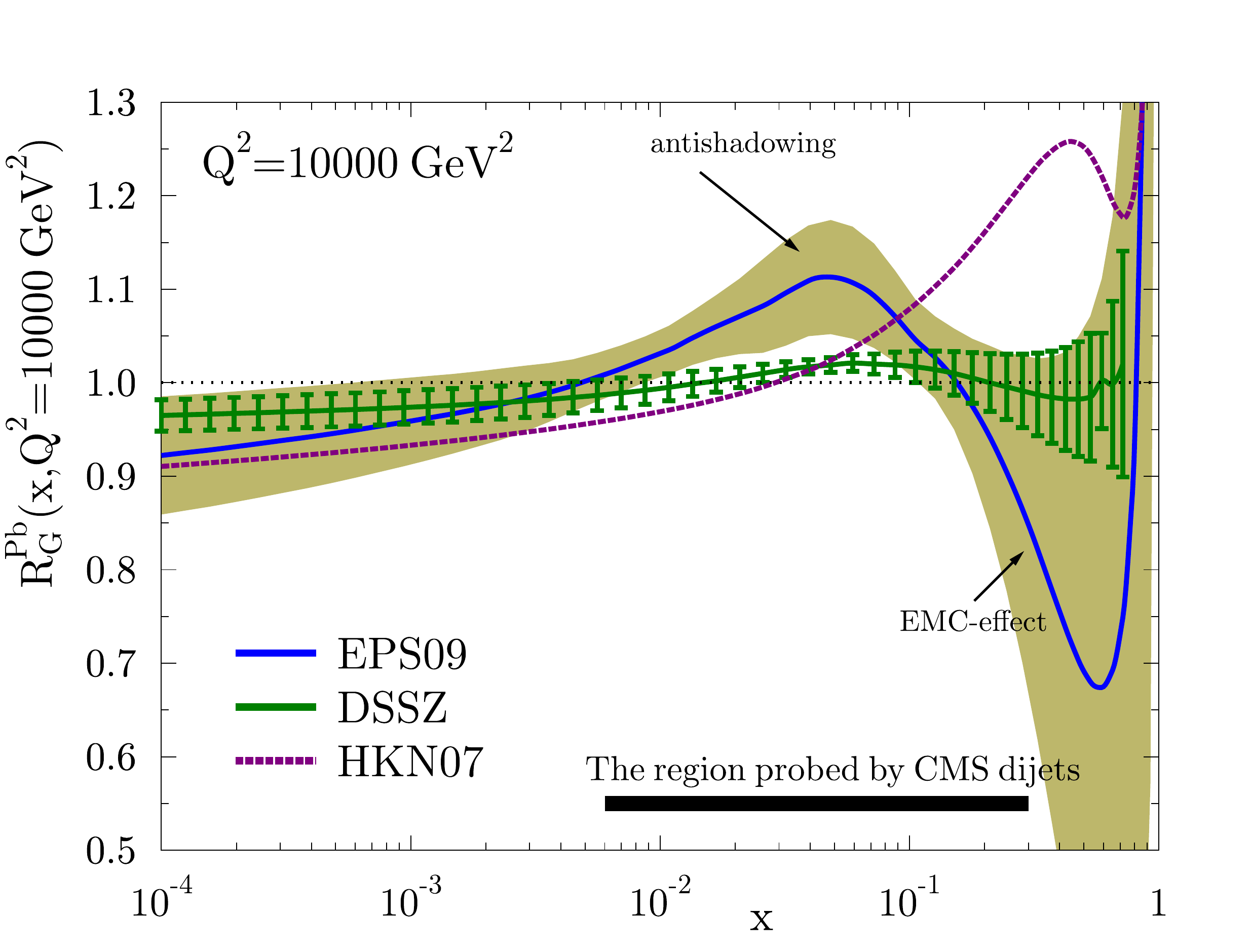}}
\caption[]{Comparison of nuclear modifications for gluon PDFs $R_{\rm G}^{\rm Pb}(x,Q) \equiv g^{\rm lead}(x,Q)/g^{\rm proton}(x,Q)$
as obtained in different fits. Figure adapted from \cite{Eskola:2013aya}.}
\label{fig:nmods}
\vspace{-0.2cm}
\end{wrapfigure}

The gluon parton distribution functions (PDFs) in heavy nuclei are not particularly well constrained \cite{Eskola:2012rg,Paukkunen:2014nqa}.
Before the nuclear collisions at the LHC, one of the very few available data directly sensitive to 
the nuclear gluons at perturbative scales were from inclusive pion production in deuteron+gold collisions
at RHIC \cite{Adler:2006wg,Abelev:2009hx}. These data were included into the EPS09 \cite{Eskola:2009uj}
global fit of nuclear PDFs (nPDFs) and gave rise to the antishadowing and EMC-effect for gluons shown in
Figure~\ref{fig:nmods} (similar results have been recently obtained by the nCTEQ collaboration \cite{Kovarik:2013sya}).
However, one can interpret the nuclear modifications seen in the RHIC pion data also as being due to nuclear
effects in the parton-to-pion fragmentation functions \cite{Sassot:2009sh} and hence reproduce the RHIC pion 
data practically without any nuclear modifications in the gluon PDFs.
This viewpoint was adopted in the DSSZ \cite{deFlorian:2011fp} global fit of nPDFs. Finally, if all the pion
data are left out, the gluons remain very weakly constrained and more fit parameters have to be fixed by hand.
An example of this kind of fit is HKN07 \cite{Hirai:2007sx}. It is this situation that the (di)jet production 
in the proton+lead (p+Pb) collisions at the LHC is expected to shed light on.

\section{Baseline uncertainties and nuclear modifications in dijets}

In this talk, we concentrate on the (minimum bias) dijets within the kinematic setup
of the measurements \cite{Chatrchyan:2014hqa} by the CMS collaboration during
the 2013 p+Pb run at the LHC. We consider the dijet distribution as a function
of dijet ``pseudorapidity'' defined as
\begin{equation}
\eta_{\rm dijet} \equiv \frac{1}{2} \left( \eta_{\rm leading} + \eta_{\rm subleading} \right), 
\end{equation}
where $\eta_{\rm leading}$ and $\eta_{\rm subleading}$ are the pseudorapidities of the two 
hardest jets in a dijet event with $p_{T, \rm leading \,\, jet} > 120 \, {\rm GeV}$, 
$p_{T, \rm leading \,\, jet} > 30 \, {\rm GeV}$, and $|\eta_{\rm leading, subleading}|<3$.
During the p+Pb run, the proton beam energy
was $E_{\rm p}=4 \, {\rm TeV}$, and the lead-ions cruised correspondingly with 
$E_{\rm Pb}= (82/208) \times 4 \, {\rm TeV} \approx 1.58 \, {\rm TeV}$. 
Because of the the different (per-nucleon) beam energies a shift between the center-of-mass and the collider
(laboratory) frame of $\eta_{\rm shift}=0.5\log(E_{\rm Pb}/E_{\rm p}) \approx -0.465$
units in rapidity is induced. As typical public Monte-Carlo jet codes (like MEKS \cite{Gao:2012he} used in this analysis)
assume that the collisions are symmetric and take place in the center-of-mass frame, some additional tweaking
is necessary to implement these asymmetric p+Pb collisions with all the experimental cuts.

\begin{figure}[th!]
\centering
\includegraphics[width=0.45\textwidth]{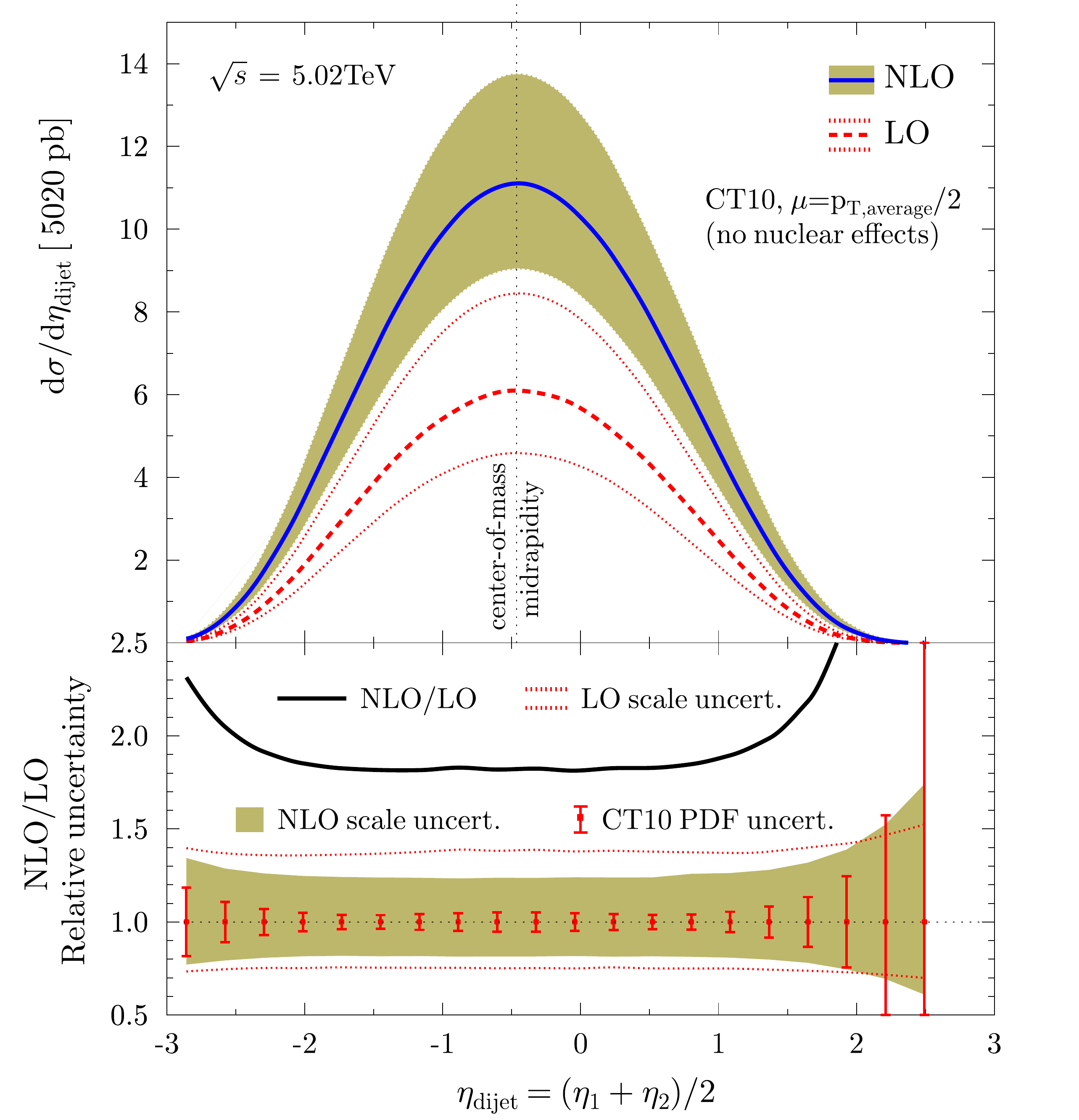}
\includegraphics[width=0.45\textwidth]{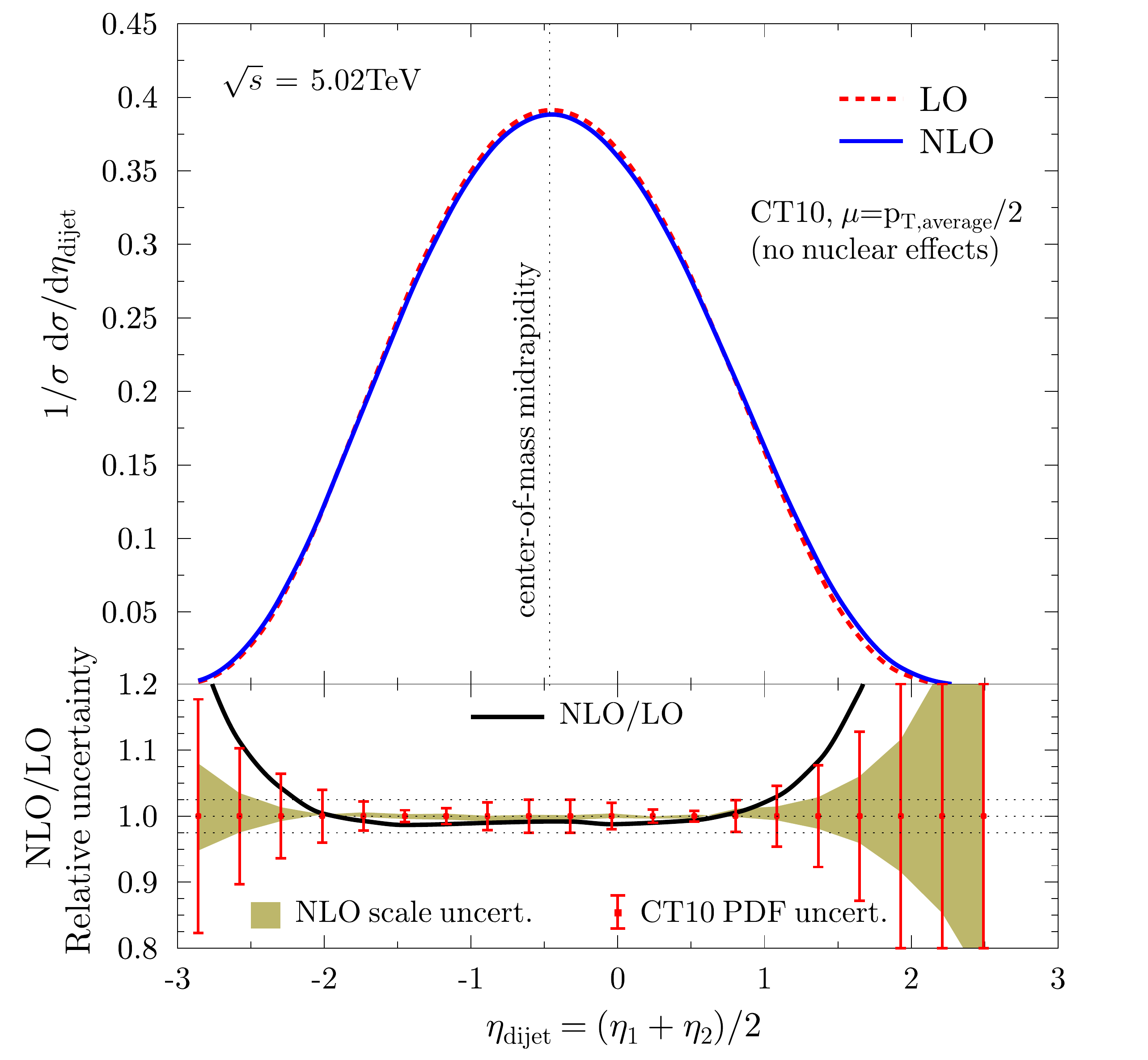}
\caption{{\bf Left-hand panel:} Absolute pseudorapidity distribution for the dijets
within the kinematic cuts of CMS.
{\bf Right-hand panel:} The distribution of the  left-hand panel normalized by
cross section integrated over the rapidity. Figures from \cite{Eskola:2013aya}.} 
\label{fig:spectra}
\end{figure}

A baseline prediction for the absolute dijet cross section is shown in Figure~\ref{fig:spectra}.
No nuclear effects are involved, but only the CT10NLO free-proton PDFs \cite{Lai:2010vv} have been used.
While the next-to-leading order (NLO) correction turns out rather large and the scale uncertainty
is significant, the crucial point is that the NLO-to-LO ``K-factor'' remains almost constant in the central region
($-2 \lesssim \eta_{\rm dijet} \lesssim 1$). 
This means that the shape of the rapidity distribution is not subject to a significant NLO correction.
This is evident in the right-hand panel where we plot the rapidity distribution normalized
by the total (rapidity-integrated) dijet cross section. Indeed, the scale uncertainty shrinks dramatically
and the CT10-originating uncertainties tend to diminish as well. Also some cancellation related to the
systematics of the underlying event can be expected to take place.

\vspace{0.1cm}

The nuclear effects in the gluon PDFs are expected to induce modifications on these
distributions differently in the proton-going and lead-going directions. Indeed, as shown in
Figure~\ref{fig:dataajasensemmoista}, the prediction with CT10 tends to underestimate
the measurements at proton-going direction and overestimate at lead-going direction.
These deficiencies are, however, readily patched by the antishadowing and EMC-effect
in the EPS09 gluon PDFs, as Figure~\ref{fig:dataajasensemmoista} demonstrates. The corresponding
calculations with DSSZ show only a little difference to the CT10 curve and the nuclear
effects from HKN07 actually correct the spectra in a wrong way.

\newpage
\section{Quantitative constraints: reweighting of EPS09}

\begin{wrapfigure}{r}{0.45\textwidth}
\vspace{-0.5cm}
\centerline{\includegraphics[width=0.5\textwidth]{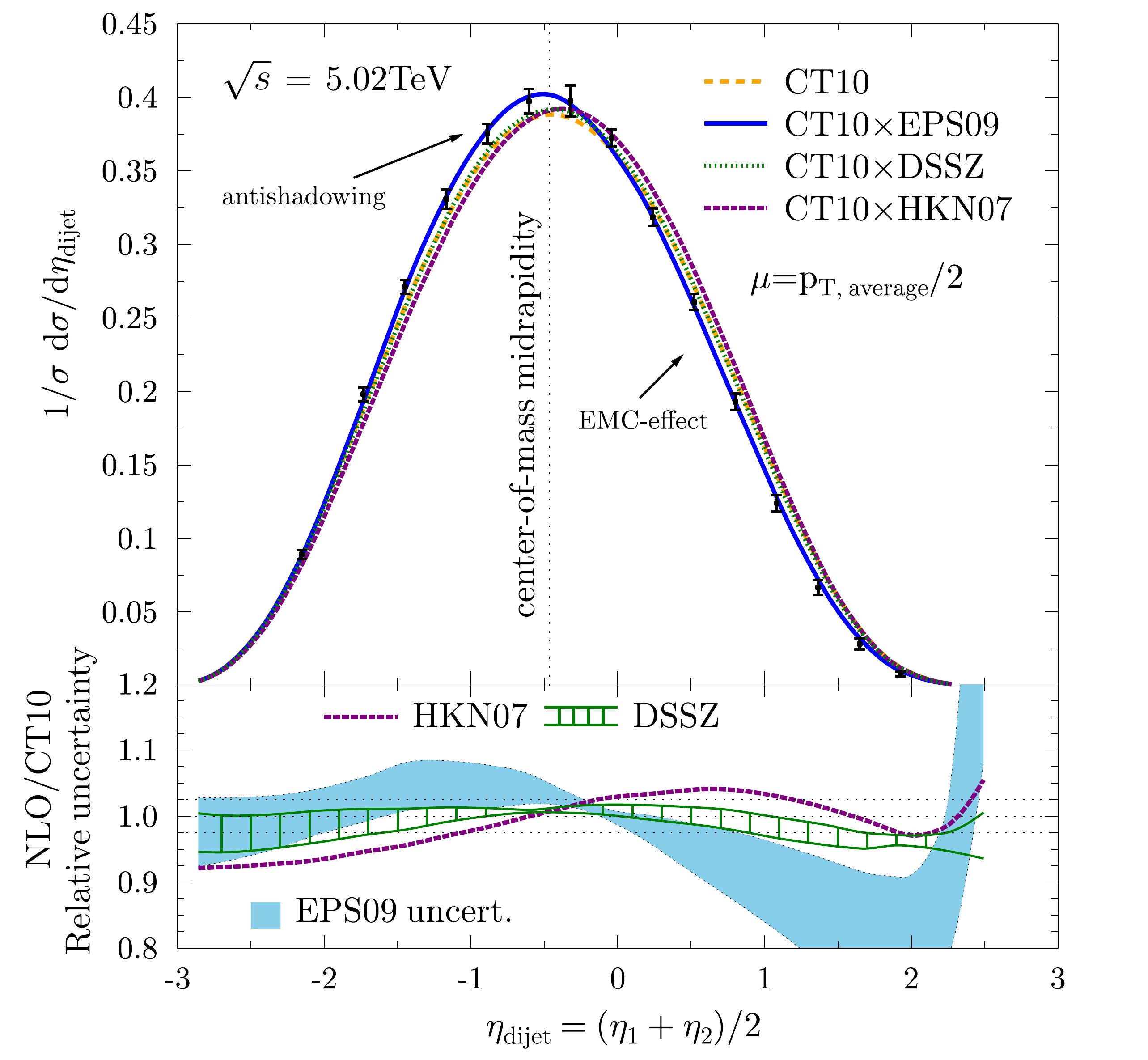}}
\caption[]{The preliminary CMS dijet data \cite{Chatrchyan:2014hqa} 
compared to predictions with different PDFs. Figure adapted from \cite{Eskola:2013aya}.}
\label{fig:dataajasensemmoista}
\end{wrapfigure}
As Figure~\ref{fig:dataajasensemmoista} already indicated, EPS09 agrees with the CMS data.
However, to better understand what kind of further constraints these data might provide,
we invoke the method of Hessian PDF reweighting \cite{Paukkunen:2013grz,Paukkunen:2014zia}:
We recall that the central set of EPS09 corresponds to a minimum of a certain global 
$\chi^2$-function which can be expanded in the vicinity of the minimum as
\vspace{-0.1cm}
\begin{equation}
 \chi^2\{a\} \approx \chi^2_0 + \sum_{ij} (a_i-a_i^0) H_{ij} (a_j-a_j^0)
 = \chi^2_0  + \sum_i  z_i^2. \label{eq:chi2orig}
\vspace{-0.1cm}
 \end{equation}
Here, $a_i$ denote the fit parameters (the best fit corresponds to $a_i=a_i^0$) and $H_{ij}$ is 
the second-derivative matrix (the Hessian matrix) which has been diagonalized in the last step.
The central PDF set $S_0$ corresponds to the origin of this
``$z$-space'' and the PDF error sets $S^\pm_k$ are defined by 
$
z_i({S^\pm_k}) =   \pm \sqrt{\Delta \chi^2} \delta_{ik},
$
where $\Delta \chi^2=50$ for EPS09. If we were to include a new set of data into our global fit,
we would naturally add its $\chi^2$-contribution on top of everything else in Eq.~(\ref{eq:chi2orig}).
Now, as the the PDF error sets are available we can realize this approximately by defining
\begin{equation}
 \chi^2_{\rm new} \equiv \chi^2_0  +   \sum_k z_k^2  + 
 \sum_{i,j} \left(y_i[f]-y_i^{\rm data}\right) C_{ij}^{-1} \left(y_j[f]-y_j^{\rm data}\right), \label{eq:newchi2} \nonumber
\end{equation}
where $y_i^{\rm data}$ are the new data points with covariance matrix $C_{ij}$. We can estimate
the theory values $y_i[f]$ linearly by
\vspace{-0.1cm}
\begin{equation}
 y_i \left[f \right] \approx y_i \left[{S_0} \right] + \sum_{k} \frac{\partial y_i [{S}]}{\partial z_k}{\Big|_{S=S_0}} z_k
                   \approx y_i \left[S_0 \right] + \sum_{k} \frac{y_i[S_k^+] - y_i[S_k^-]}{2} \frac{z_k}{\sqrt{\Delta \chi^2}}, \label{eq:XS}
\vspace{-0.1cm}
\end{equation}
and, in this way, $\chi^2_{\rm new}$ becomes a quadratic function of the variables $z_i$ and it has
a well-defined minimum denoted here by $z_i=z^{\rm min}_k$. The corresponding set of PDFs $f_i^{\rm new}(x,Q^2)$ can be computed by
\vspace{-0.1cm}
\begin{equation}
 f_i^{\rm new}(x,Q^2) \approx f_i^{S_0}(x,Q^2) + \sum_{k}  \frac{f^{S^+_k}_i(x,Q^2)-f_i^{S^-_k}(x,Q^2)}{2}  \frac{z^{\rm min}_k}{\sqrt{\Delta \chi^2}}. \label{eq:newPDF}
 \vspace{-0.1cm}
\end{equation}
After finding the minimum, one can also construct the new error sets similarly as sketched above.
\begin{figure}[th!]
\vspace{-0.2cm}
\centering
\includegraphics[width=0.40\textwidth]{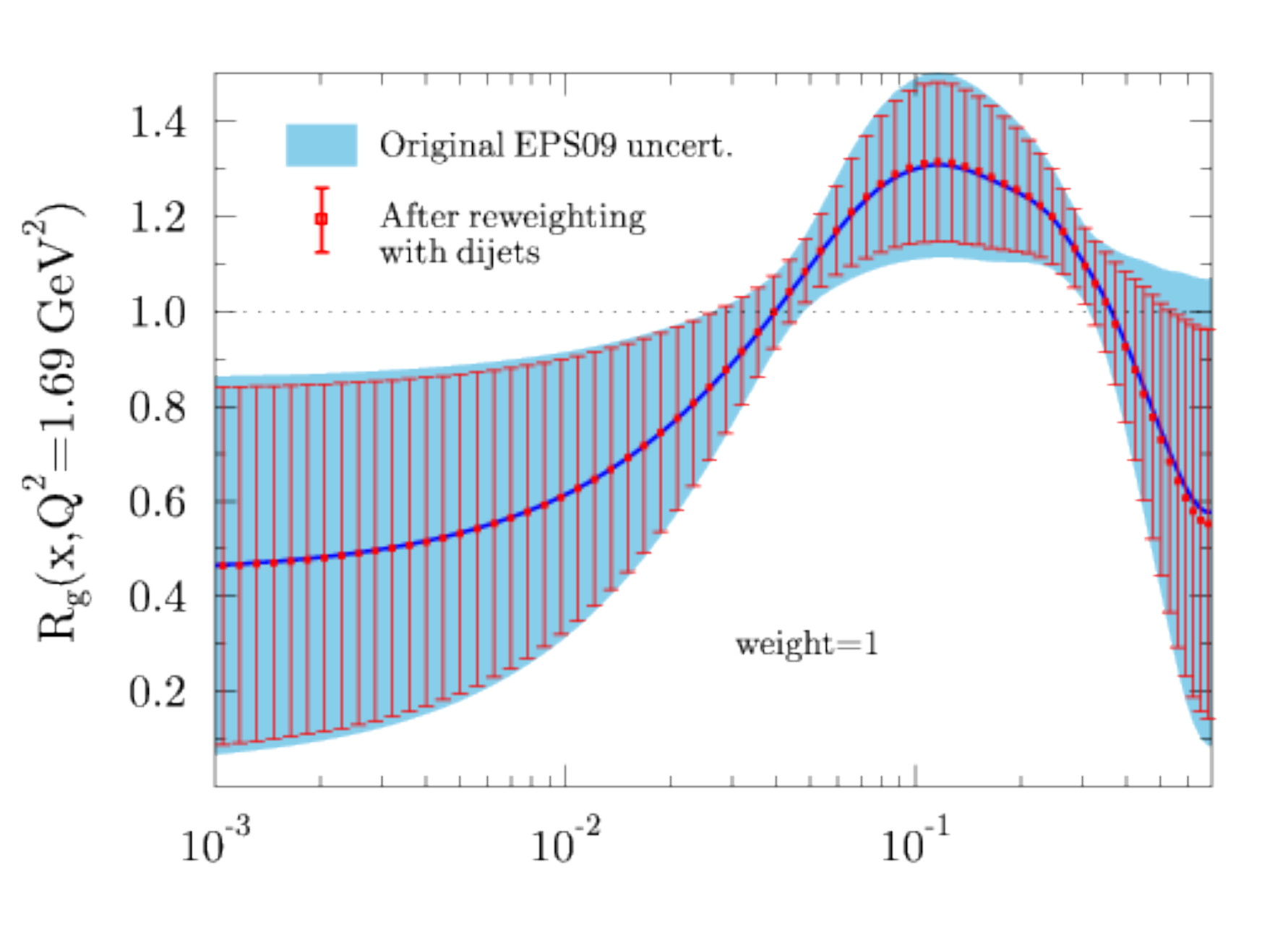}
\includegraphics[width=0.40\textwidth]{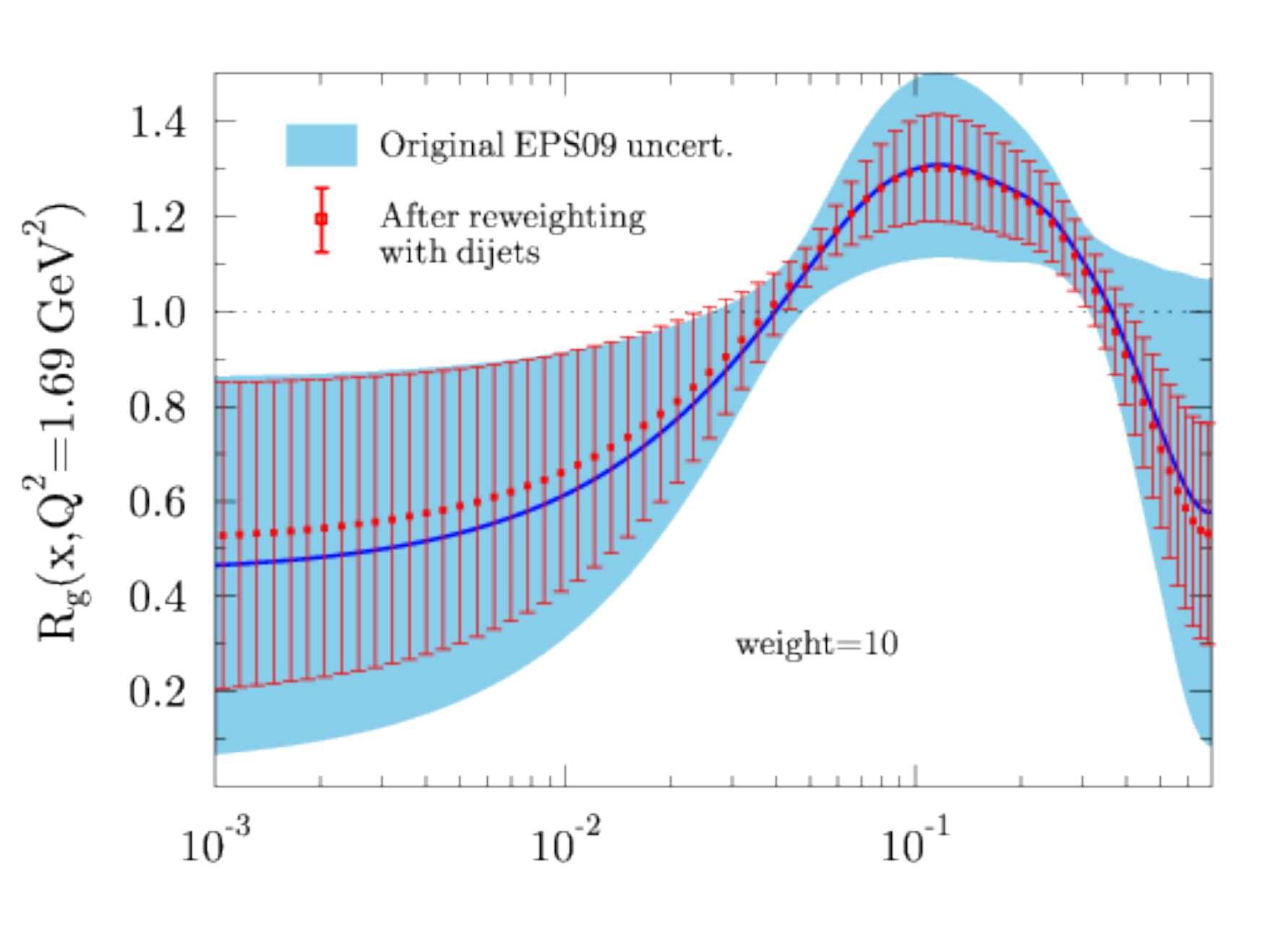}
\caption{ {\bf Left-hand panel:} The EPS09 nuclear modification $R_G(x,Q^2=1.69 \, {\rm GeV}^2)$ 
before and after the reweighting with CMS p+Pb dijet data. {\bf Right-hand panel:} As the left-hand panel
but giving the dijet data an extra weight of 10.} 
\label{fig:rw}
\vspace{-0.2cm}
\end{figure}

\newpage
The effect of reweighting EPS09 with the dijet data shown in Figure~\ref{fig:dataajasensemmoista} 
is presented in Figure~\ref{fig:rw}. Here, we have taken all the data uncertainties as uncorrelated
and treated the CT10 uncertainties as correlated systematic uncertainties when forming the covariance
matrix. As all the members of the
EPS09 package already have the gluon antishadowing built in, the effect of reweighting is not
dramatic (left-hand panel). To better appreciate what the data implies we have repeated the
reweighting giving these dijet data an additional weight of 10 (right-hand panel). This exercise
not only shows that these data will mostly affect the large-$x$ gluons, but demonstrates that
they are also in a perfect agreement with EPS09 (the new error band is inside the old one).

\vspace{-0.2cm}
\section{Forward-to-backward asymmetry}

\begin{wrapfigure}{r}{0.37\textwidth}
\vspace{-1.0cm}
\centerline{\includegraphics[width=0.40\textwidth]{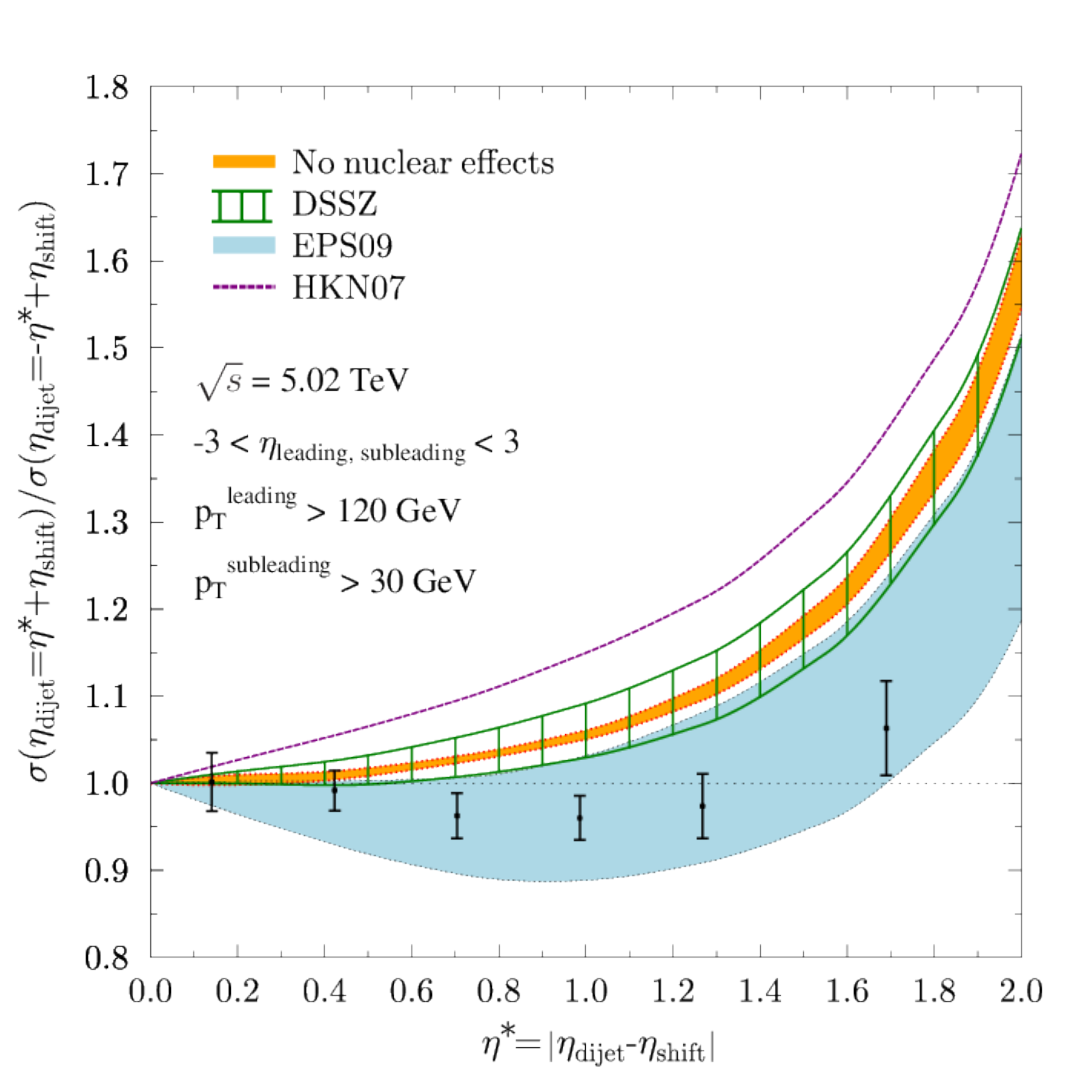}}
\caption[]{Dijet forward-to-backward asymmetry. Data points have been
estimated from \cite{Chatrchyan:2014hqa} by adding all uncertainties
in quadrature. Figure adapted from \cite{Eskola:2013aya}.}
\label{fig:fb}
\vspace{-1.0cm}
\end{wrapfigure}
To highlight the nuclear effects in the dijet spectrum, a better
option is to form the forward-to-backward ratio by dividing the yields
in the forward direction by the corresponding yields in the backward direction,
symmetrically around the center-of-mass midrapidity.
Figure~\ref{fig:fb} presents our predictions using different PDFs together
with the CMS data points (constructed by hand from those shown in Figure~\ref{fig:dataajasensemmoista}).
Because of the asymmetric detector cuts in the center-of-mass frame, there is more
open phase space in the forward direction and the cross sections tend to be larger there
in comparison to the backward direction. For this reason even the prediction with
no nuclear effect (just CT10) is above unity. Nevertheless, the CMS data clearly deviate
from the baseline prediction (even more from HKN07) and are in good agreement with EPS09.

\vspace{-0.2cm}
\section{Conclusions}

In summary, we have discussed the dijet production in p+Pb collisions with a special
focus on the recent measurements by the CMS collaboration. The normalized rapidity distribution
around the midrapidity is surprisingly well protected against higher-order
QCD corrections and in this sense the agreement we find using the EPS09 nuclear PDFs appears solid.
The PDF-reweighting studies indicate that these data should offer novel constraints
for the nuclear gluon PDFs at large $x$ where the uncertainty is currently large.

\vspace{-0.2cm}
\section*{Acknowledgments}
H.~P. and K.~J.~E. acknowledge the financial support from the Academy of Finland, Project No. 133005. 
C.~A.~S. is supported by European Research Council grant HotLHC ERC-2011-StG-279579, and by Xunta de Galicia.

\end{document}